# Effects of fabrication routes and material parameters on the control of superconducting currents by gate voltage

L. Ruf[1], T. Elalaily[2,3,4], C. Puglia[5], Yu. P. Ivanov[6], F. Joint[7], M. Berke[2,3], A. Iorio[5], P. Makk[2,3], G. De Simoni[5], S. Gasparinetti[7], G. Divitini[6], S. Csonka[2,3], F. Giazotto[5], E. Scheer[1†], A. Di Bernardo[1,8*]

1. *Department of Physics, University of Konstanz, 78457 Konstanz, Germany.*
2. *Department of Physics, Budapest University for Technology and Economics, 1111 Budapest, Hungary.*
3. *MTA-BME Superconducting Nanoelectronics Momentum Research Group, Budapest, Hungary*
4. *Department of Physics, Faculty of Science, Tanta University, Al-Geish St., 31527 Tanta, Gharbia, Egypt*
5. *NEST, Istituto Nanoscienze and Scuola Normale Superiore, 56127 Pisa, Italy.*
6. *Istituto Italiano di Tecnologia, 16163 Genova, Italy.*
7. *Department of Microtechnology and Nanoscience, Chalmers University of Technology, 41296 Göteborg, Sweden.*
8. *Department of Physics "E. R. Caianiello", Università degli studi di Salerno, 84084 Fisciano, SA, Italy.*

[†]Email: elke.scheer@uni-konstanz.de
[*]Email: angelo.dibernardo@uni-konstanz.de

**Abstract**

The control of a superconducting current via the application of a gate voltage has been recently demonstrated in a variety of superconducting devices. Although the mechanism underlying this gate-controlled supercurrent (GCS) effect remains under debate, the GCS effect has raised great interest for the development of the superconducting equivalent of conventional metal-oxide semiconductor electronics. To date, however, the GCS effect has been mostly observed in superconducting devices made by additive patterning. Here, we show that devices made by subtractive patterning show a systematic absence of the GCS effect. Doing a microstructural analysis of these devices and comparing them to devices made by additive patterning, where we observe a GCS, we identify some material and physical parameters that are crucial for the observation of a GCS. We also show that some of the mechanisms proposed to explain the origin of the GCS effect are not universally relevant.

## 1. Introduction

In conventional metal-oxide semiconductor (CMOS) circuits, the logic state of one of the elemental three-terminal device components (the transistor), is controlled via the application of a gate voltage ($V_G$). The applied $V_G$ induces an electric field ($E$) that changes the density of charge carriers flowing through a nanoscale-size constriction of the transistor and this in turn sets the logic state of the device.

The superconducting equivalent of such effect had remained unknown for years, possibly because it was believed that in a superconductor (S), which is a normal metal (N) above its superconducting critical temperature ($T_c$), the $E$ induced by an applied $V_G$ would be just screened within the Thomas-Fermi length[1-2] (typically a few angstroms from the S surface[3]), meaning that $E$ would have no effects on the S properties.

Over the past few years, however, several groups[4-22] have shown that an applied $V_G$ can affect the superconducting current (supercurrent) through a nanoconstriction made from a S. As the applied $V_G$ is increased, the critical supercurrent ($I_c$) of the S nanoconstriction does not change significantly compared to its value measured at $V_G = 0$, until after $|V_G|$ reaches a certain threshold value ($V_{G,onset}$). For $|V_G| > V_{G,onset}$, $I_c$ gets progressively suppressed until it becomes null at an even higher $|V_G|$ ($V_{G,offset}$). The applied $V_G$ can therefore switch the S nanoconstriction from a superconducting state with zero resistance (and $I_c \neq 0$) to a metallic state with non-null resistance (and $I_c = 0$).

The possibility of switching a superconducting device between two states with different resistance via an applied $V_G$ can be seen as the superconducting equivalent of the effect used to control the logic state of semiconductor transistors in CMOS electronics. This phenomenon, which we name gate-controlled supercurrent (GCS) effect as in Ref. 20, is fully reversible, as it has been shown that superconducting devices exhibiting a GCS effect can be freely switched between the resistive and superconducting states, upon the application and removal of $V_{G,offset}$, respectively [4,9-13,15-17,19-20]. Other physical features of the GCS effect confirmed by the majority of the studies done to date include that the effect is independent of the $V_G$ polarity[4-9,11,12,19,20], it decays over a length scale of the order of the S coherence length[4,15], and it is weakly dependent on temperature ($T$) and applied magnetic field ($H$)[4,5,8,9,11,12,15,16,19,20,22], meaning that $V_{G,\text{offset}}$ does not change significantly with $T$ and $H$.

Although the experimental signatures of the GCS effect are well-established, the physical mechanism responsible for the effect is still under debate. Understanding the main mechanisms at play for the GCS remains a matter of priority to control the effect, which is in turn crucial for the development of future applications based on it.

Several mechanisms have been proposed to explain the origin of the GCS effect. Also, it should be noted that, in some studies, and in earlier ones, the authors do not specify exactly the mechanism at play in their experiment. Some research groups have argued that high-energy electrons emitted from the gate electrode into the S can excite phonons and/or decay into quasiparticles in the S thus suppressing $I_c$ (scenario 1)[15,17,18]. Other studies suggest that the leakage current ($I_{\text{leak}}$), which flows from the gate electrode into the S nanoconstriction upon the $V_G$ application, can induce heating of the electronic system due to the phonons triggered in the substrate (scenario 2)[15,16,19,23] or drive the S into an out-of-equilibrium state with phase fluctuations but without sizable heating (scenario 3)[16,20,21]. Last, several other groups ascribe the GCS effect to an effect induced by the $E$ associated to the applied $V_G$ (scenario 4)[4,14,22].

Independently on what the specific mechanism underlying the GCS effect is, we note that to date a GCS has been observed in superconducting devices based on a variety of Ss (e.g., Al[4,9,19,23], Nb[6,8,15], V[10,18], Ti[4,6,7,11,15-17], W-C[22], Ta[20], TiN[15]) and with different geometries including nanowires[4,14-17,19,20], Dayem bridges[7-11,14], S/N/S Josephson junctions (with $V_G$ applied to the N weak link, N being a normal metal)[5,21], superconducting interferometers[6,13] and resonators[18,23]. For superconducting resonator devices , the resonance frequency $f_0$ other than $I_c$ is the physical parameter the variation of which is tracked upon the application of a $V_G$[18,23].

Most of the devices, however, where the GCS effect has been reported have been fabricated using a bottom-up fabrication route [4,13,17,19-21,23] (i.e., by additive patterning). Here, we show that devices made following a top-down approach (i.e., by subtractive patterning) show a systematic absence of the GCS effect, independently on their geometry and on the S used for their fabrication. Given the absence of the GCS effect, our gate-controlled superconducting devices made by subtractive patterning represent an ideal system to determine the parameters which are responsible for the absence of the GCS effect and to discuss them in the light of the mechanisms proposed in the literature.

Performing a microstructural characterization of the devices made by subtractive patterning with no GCS and comparing them to devices made by additive patterning, which instead show a GCS, we identify some material parameters that are different between the two types of devices, and which represent therefore key factors for the GCS observation. Our analysis also suggests that some of the mechanisms proposed in the literature to explain the GCS effect cannot account for the different behavior of devices made by subtractive or additive patterning.

## 2. Experimental

*Sample fabrication* - We have made superconducting devices with gate electrodes by both subtractive and additive patterning, to which we refer as etched devices and lift-off devices, respectively, since their corresponding fabrication process involves an etching step or a lift-off step (see section 3). To minimize hidden parameters in the sample fabrication as origin of the different behavior regarding the CGS effect, we have fabricated devices in different geometries (Dayem bridges and nanowires) different materials in two different labs each. Nb Dayem bridge devices have been fabricated by two of our groups, at the University of Konstanz (UKON) and at the Centro Nazionale delle Ricerche (CNR), following protocols involving both subtractive and additive patterning. The gate-controlled Nb devices made by lift-off at CNR have been fabricated following the process described in Ref. 8. The detailed fabrication procedure and parameters for the devices made at UKON as well as for the etched devices at CNR are given in the Supplementary Material.

Gate-controlled NbTiN devices with a Dayem bridge geometry made by both lift-off and etching have been fabricated at the Budapest University of Technology and Economic (BME). In addition, NbTiN etched devices with a nanowire geometry have been realized at the Chalmers University of Technology (CUT). Also for these devices the detailed fabrication protocols are given in the Supplementary Material.

*Transport measurements* - The current-voltage characteristics (*IV*s) of the devices have been measured in the labs where the respective samples have been fabricated by sweeping a current and measuring the voltage drop. The etched and lift-off samples fabricated in the same lab have been studied in the same cryostat using the same wiring to avoid possible impacts of the measurement setups or routines. The *IV*s have always been recorded for both sweep directions. Within the intrinsic variation of switching current distributions typical for such devices[20-21], the *IV*s are mirror symmetric upon reversal of the sweep direction as typical for hysteretic Josephson junctions. Therefore, for clarity, we show in this work always *IV*s recorded for one sweep direction, namely for increasing bias current.

## 3. Results and discussion

### A. Fabrication routes of gate-controlled superconducting devices

To better understand the mechanisms underlying the GCS effect and the material/device parameters controlling it, we have fabricated a series of devices based on different Ss and fabrication processes. The fabrication recipes, given in detail in the Supplementary Material, are essentially of two different types. The first type of fabrication route is shown in Fig. 1a. This is a top-down fabrication based on subtractive patterning which starts with the deposition of a S thin film onto an insulating substrate. Once the S thin film is grown, a negative resist is spin-coated onto it, which is then patterned by EBL into the desired device geometry. After development of the unexposed resist, the patterned resist is used as a mask during the following etching process which transfers the device pattern into the S (Fig. 1a). Last, the resist mask is removed leaving the desired device. Due to the etching step involved in the fabrication, we refer to gate-controlled devices fabricated with this top-down approach also as etched devices.

The second type of fabrication process shown in Fig. 1b is a bottom-up approach based on additive patterning, where the gate-controlled device is patterned EBL into a positive resist, after this is spun onto an insulating substrate. After EBL patterning, the resist is developed and then the S material is deposited (usually by sputtering or evaporation). Last, the resist is removed with a solvent (lift-off step) which leaves the desired superconducting device. Due to this last lift-off step, we also refer to devices made with this bottom-up approach as lift-off devices.

We note that dry etching has already been used by a few other groups for the fabrication of gate-controlled superconducting devices[14-16,18,24]. Most of these devices[15,16], however, have been made onto a Si substrate without an insulating $SiO_2$ layer. The GCS effect in these devices seems mostly dominated by $I_{leak}$-induced dissipation due the stronger thermal coupling between the S nanoconstriction and the substrate due to the absence of an insulating layer (scenario 2 above). This is evidenced by the fact that the typical $E$ corresponding to $V_{G,offset}$ is much lower (~ 0.5 MV/cm; Refs. 15,16) than that reported for devices made on an insulating substrate (~ 4 MV/cm; Refs. 4-7,17). In two other reports, where etching has been used and the substrate is insulating, either a limited suppression of the superconducting state[14,18] or even an increase in $I_c$ (Ref. 24) under the applied $V_G$ has been instead reported.

Across our research groups, we have fabricated a variety of gate-controlled devices by dry etching. We have used Nb and NbTiN on various insulating substrates (300-nm-thick $SiO_2$ on p-doped or intrinsic Si, and $Al_2O_3$), as reported in the Supplementary Material and shown in Fig. 2 and in the Supplementary Figs. S1 to S3. Unlike lift-off devices made of the same S material and with the same geometry, for which we observe a GCS effect, all these etched devices exhibit no GCS.

## B. Gate-controlled superconducting current effect in etched and lift-off devices

Fig. 2 shows an example of a lift-off and of an etched gate-controlled nanowire made of the same S (i.e., NbTiN). To show the effect of $V_G$ on $I_c$, for each device we report a few representative current versus voltage characteristics (IVs) as a function of the applied $V_G$. Although both nanowires are superconducting with critical temperature $T_c \sim 12.5$ K (Fig. 2c and Fig. S1), we find that in the etched device (Fig. 2a), both the $I_c$ and the $I_r$ ($I_r$ being the retrapping current) are completely unaffected by the applied $V_G$ (Fig. 2d), which demonstrates that the etched device shows no GCS. We observe the absence of a GCS effect in these NbTiN etched devices not only when $V_G$ is applied through a side gate (Figs. 2b and 2d), but also when $V_G$ is applied to the $SiO_2$/p-doped Si used as back gate (Fig. S2). Also, the power which is dissipated by the gate $P_G = V_G I_{leak}$ at the largest applied $V_G = 120$ V for these etched devices is comparable to the power $P_N = R_N I_r^2$ that the device would dissipate when it switches to the normal state ($R_N$ being the normal-state resistance of the device). This consideration suggests that, despite phonon-induced heating associated with $I_{leak}$ can be significant in these devices, no GCS effect is observed. When gate-controlled devices based on the same S (NbTiN) devices are fabricated by lift-off instead, we can observe a full GCS effect, as evidenced by the *IV* curves at different applied $V_G$ values in Fig. 2e – which we have measured for the device shown in Fig. 2b.

## C. Microstructural characterization of devices and analysis of mechanisms responsible for their different behavior

As reported in the Supplementary Material, we have fabricated and tested almost 30 etched devices based on the above-listed S materials, and also having different geometries (i.e., both Dayem bridges and nanowires). In Fig. S3, we show the absence of the GCS in etched NbTiN nanowires with a different geometry than those reported in Figs. 2 and S1, whilst in Fig. S4 we show the absence of the GCS in etched Nb Dayem bridges.

All the etched devices which we have made and tested, except for one with very high $I_{leak}$, do not show a GCS effect, even for $I_{leak}$ up to several tens of nA and applied $V_G$ up to or above 100 V (see Supplementary Material). We outline that the GCS effect is absent in all these devices, even though they have been made with the identical geometry used for lift-off devices for which we instead observe a GCS effect. This observation suggests that geometry is most likely not a factor that plays a key role towards a GCS.

Keeping in mind the fabrication steps for etched and lift-off devices (illustrated in Fig. 1) and assuming that the mechanism responsible for the GCS is one of those proposed in the literature (scenarios 1 to 4 listed above), we argue that the following differences may be responsible for the different behavior of the two types of devices with respect to the GCS:

A) If field emission of hot electrons is responsible for the $I_c$ suppression (scenario 1), the redeposition of oxide species from the substrate onto the walls of the S constriction during the etching step can make the tunneling of hot electrons into the S less efficient for etched devices. The reasoning behind this argument that a thicker oxide layer on the surface of the S nanoconstriction may stop electrons emitted from the gate more effectively than a thinner one, and thereby reduce their impact onto the superconducting state.

B) If phonon heating is responsible for the $I_c$ suppression (scenario 2), then physical etching into the substrate should increase the propagation length of phonons that reach the device and, as a consequence, suppress the GCS in etched devices.

C) If $I_{leak}$-induced phase fluctuations (scenario 3) or an $E$-driven effect (scenario 4) are the mechanisms responsible for the suppression of $I_c$, then differences in the microstructure of the S material (e.g., grain size, presence of dislocations), or in the S surface can account for the absence of the GCS in etched devices. Structural parameters such as grain size, shape or roughness of the S can in principle be different for lift-off or etched devices. Also, an $E$-effect or $I_{leak}$-induced phase fluctuations can be enhanced by surface states in the S, which in turn can change depending on the S surface morphology. The etched process can also introduce changes in the S surface[25,26], which can explain the different behavior of etched and lift-off devices.

To rule out or validate some of the above hypotheses, we have carried out scanning transmission electron microscopy (STEM) imaging and electron-energy loss spectroscopy (EELS) analysis of lamellae fabricated from etched devices of Nb showing no GCS effect and from lift-off devices of Nb showing a GCS effect. The results of our STEM and EELS analysis are shown in Figs. 3a to 3d for an etched Nb device and in Figs. 3e to 3h for a lift-off Nb device.

The EELS analysis shows that nanowires made by etching have $Nb_2O_5$ layers on their side edges which are thinner than those of devices made by lift-off, where the contact between Nb and the EBL resist before lift-off possibly results in the formation of a thicker $Nb_2O_5$ layer. This observation suggests that the disruption of superconductivity due to high-energy electrons tunneling into the S should be even more efficient in etched devices compared to lift-off devices, which rules out the case A listed above and hence scenario 1) as general mechanism responsible for the GCS effect.

The lamellae fabricated on Nb etched devices show that we only etch a few nanometers deep into the $SiO_2$ substrate, meaning that the absence of a GCS in etched devices cannot be ascribed to an increase in the propagation distance for phonons compared to lift-off devices in our devices. This observation rules out the case B above and therefore scenario 2 as main mechanism behind the GCS effect in our devices.

The images and the STEM-EELS elemental maps in Figs. 3d and 3h display structural differences: lift-off devices have a rougher interface between Ti and the $SiO_2$ substrate compared to etched devices, possibly due to polymeric residues that are left in the trenches of the patterned resist after its development and before the deposition of the S material. This increase in surface roughness for nanowires made by lift-off can lead to an enhancement of the $E$ at the nanowire/substrate interface in lift-off devices, which is consistent with the scenario C discussed above, meaning with enhanced $I_{leak}$-induced phase fluctuations or $E$-induced effects.

Also, we observe that Nb nanowires made by lift-off show significant bending on the edges (Figs. 3f and 3g) possibly induced by the mechanical pulling force that the resist exerts on the wire during the lift-off process. As for the increase in interface roughness in lift-off devices compared to etched devices, the presence of bending in lift-off devices can also cause variations in microstrain and in turn an enhancement in the local $E$ gradient. This can be another reason why a GCS is usually observed in lift-off devices, but not in etched devices.

Surface changes induced by the fabrication process can also account for the different behavior of etched and lift-off devices with respect to the GCS effect. Spectroscopy measurements with surface-sensitive techniques like nano angle-resolved photoemission spectroscopy (nano-ARPES) under an applied $V_G$ can be used in the future to study the evolution of surface states in etched and lift-off devices and confirm the relevance of surface states for the GCS effect.

Our results will certainly stimulate future studies, where structural parameters like disorder and surface roughness are systematically varied, for example by changing growth conditions of the S or using S with smaller grain size and lower crystallinity, to determine their optimal values for the GCS and to achieve a reduction in the $V_G$ needed for a full $I_c$ suppression. This systematic investigation can possibly lead to the determination of material parameters that are suitable also for the realization of the GCS in etched devices.

Achieving a reproducibility of the GCS effect in etched superconducting devices would pave the way for their integration in more complex superconducting logic circuits, since etched devices are easier to scale up compared to lift-off devices.

## 4. Conclusions

We have shown the systematic absence of the GCS effect in gate-controlled superconducting devices done by dry etching and used them as platform to study the reasons behind the absence of the GCS effect. To this aim, we have performed a microstructural characterization of the same devices in comparison to lift-off devices based on the same geometry and S materials, for which we instead observe the GCS effect.

We find that lift-off devices show a rougher and more disordered interface between the S and the substrate as well as bending towards the edges compared to etched devices. We conclude that change in these material parameters (i.e., roughness, disorder, microstrain due to bending) at the boundaries of the S constriction can affect surface states in the S and change its response to an applied $V_G$.

By comparing the behavior of etched and lift-off devices, we also show that some of the mechanisms proposed to date to explain the GCS (i.e., high-energy electron tunneling or heating due to $I_{leak}$) cannot account for the absence of the GCS in etched devices, and therefore are possibly not universal mechanisms underlying the GCS effect.

**Supplementary Material**

See the supplementary material for further details on sample fabrication, for statistics on the GCS observation in etched and lift-off devices, and for Supplementary Figures with additional experimental data on etched and lift-off devices.

**Acknowledgments**

We acknowledge support from the European Union's Horizon 2020 research and innovation program under Grant Agreement No. 964398 (SuperGate).

**Author declarations**

**Conflict of interest**

The authors have no conflicts to declare.

**Author contributions**

**Leon Ruf**: Investigation (equal); Data curation (equal); Writing – original draft (supporting). **Tosson Elalaily**: Investigation (equal); Data curation (equal); Writing – review and editing (supporting). **Claudio Puglia**: Investigation (equal); Data curation (equal); Writing – review and editing (supporting). **Yurii P. Ivanov**: Investigation (equal); Writing – review and editing (supporting); **Francois Joint**: Investigation (equal); Data curation (equal). **Martin Berke**: Investigation (equal). **Andrea Iorio**: Investigation (supporting). **Peter Makk**: Investigation (supporting); Writing – review and editing (supporting). **Giorgio De Simoni**: Investigation (supporting); Writing – review and editing (supporting). **Simone Gasparinetti**: Funding acquisition (equal); Resources (equal); Writing – review and editing (supporting). **Giorgio Divitini**: Resources (supporting); Writing – review and editing (supporting). **Szabolcs

**Csonka**: Funding acquisition (equal); Resources (equal); Writing – review and editing (supporting). **Francesco Giazotto**: Funding acquisition (equal); Resources (equal); Writing – review and editing (supporting). **Elke Scheer**: Supervision (equal); Funding acquisition (equal); Resources (equal); Writing – review and editing (supporting). **Angelo Di Bernardo**: Conceptualization (lead); Data curation (equal); Supervision (equal); Funding acquisition (equal); Resources (equal); Writing – original draft, review, and editing (lead).

## Data availability

Data will be made available from the corresponding authors upon request.

**Figures with captions**

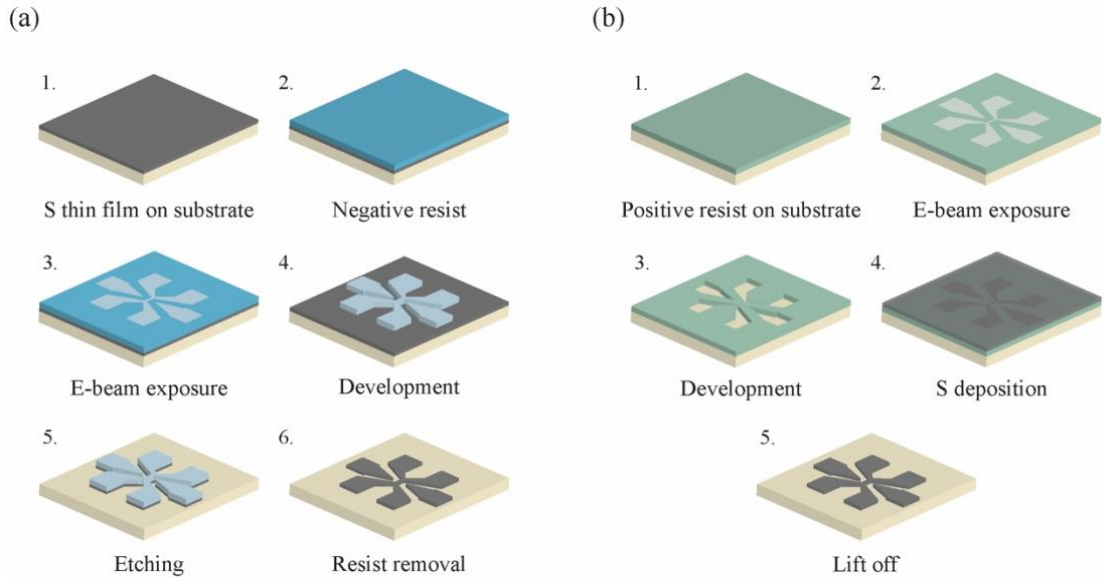

**Fig. 1**. Illustration of fabrication steps (in progressive order as specified by corresponding numbers) for the realization of gate-controlled superconducting devices with subtractive patterning (a) and additive patterning (b).

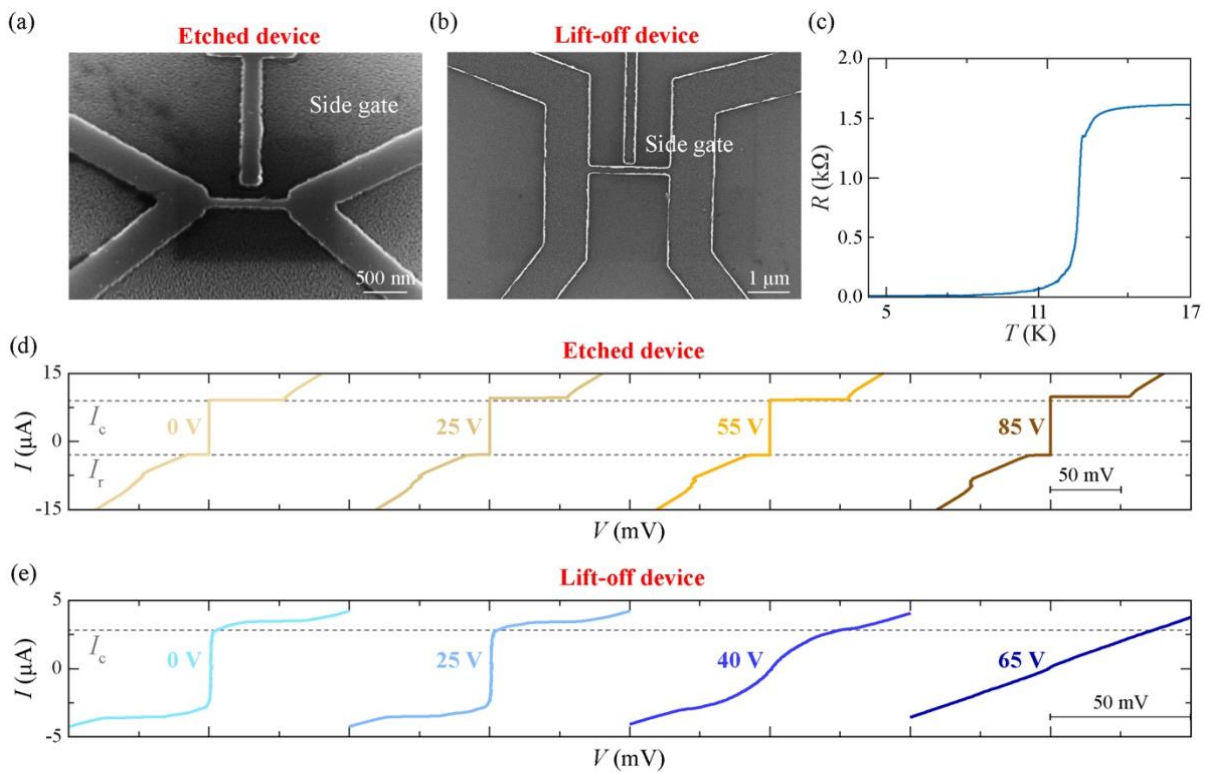

**Fig. 2.** Characterization of gate-controlled NbTiN devices made by dry etching and lift-off. (a-b) Scanning electron microscope images of a NbTiN nanowire devices made by dry etching

(a) and by lift-off (b) on a SiO$_2$ (300 nm)/*p*-doped Si substrate. (c) Resistance versus temperature, *R*(*T*) curve close around the superconducting transition for the device shown in (a). (d-e) Current versus voltage, *IV*, characteristics measured for increasing bias current *I* for the NbTiN device in (a) are shown in panel (d), and *IV* characteristics for the NbTiN device in (b) are shown in panel (e) for a few representative applied $V_G$ values (indicated next to the corresponding *IV* curve). The data in (d) for the etched device do not show a progressive suppression of either the critical current ($I_c$) or retrapping current ($I_r$) with increasing $V_G$, whilst $I_c$ is instead suppressed for the lift-off device in (e).

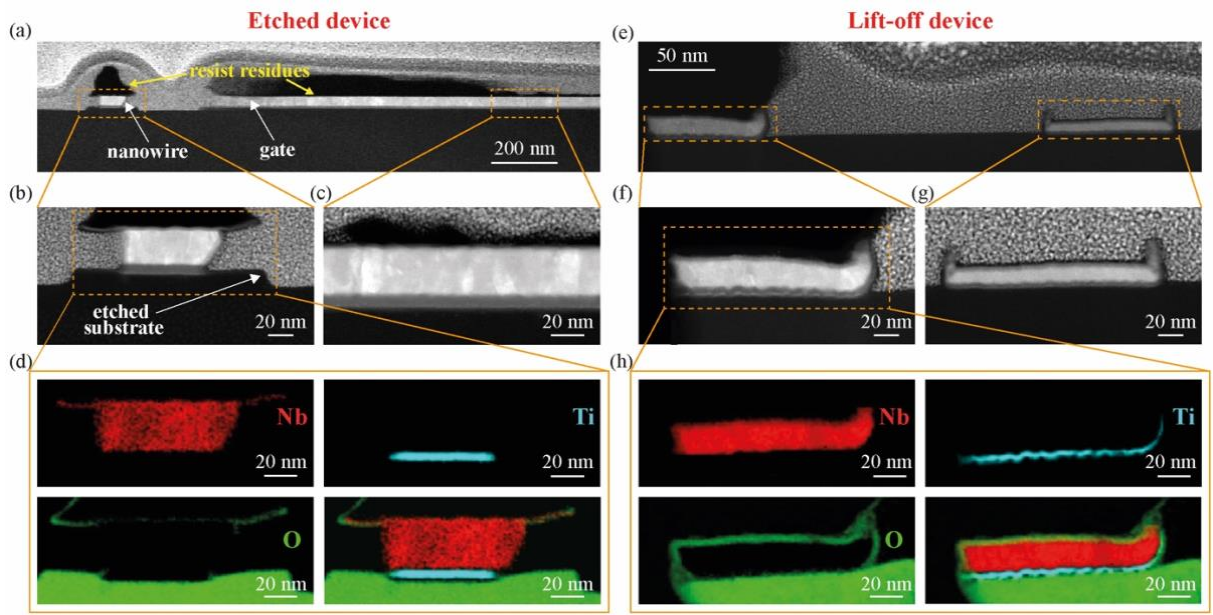

**Fig. 3.** Microstructural characterization and compositional analysis of lift-off and etched nanowires. Scanning transmission electron images (STEM) at low magnification of a Nb nanowire with side gate made by etching (a) and STEM images at higher magnification of the same nanowire (b) and gate (c) corresponding to the areas in the yellow boxes of panel (a). Electron energy loss spectroscopy (EELS) elemental maps corresponding to the area in yellow in (b) are displayed for Nb (red), Ti (light blue), O (green) and in a composite image (d). STEM images and EELS maps corresponding to those in (a-d) but obtained for a Nb lift-off device

are shown in panels (e-h) with STEM at lower magnification in (e) and at higher magnification in (f-g), and EELS maps corresponding to the area in the yellow box in (f) shown for Nb, Ti, O and all elements combined in (h).